\documentclass[prd,aps,nofootinbib]{revtex4}
\usepackage{amsmath}
\usepackage{graphicx}
\usepackage{amsmath}
\usepackage{amssymb}
\usepackage{latexsym}
\usepackage{color}

\def\be{\begin{equation}}
\def\ee{\end{equation}}
\newcommand{\swa}{Schwarzschild}
\newcommand{\ve}{\varepsilon}
\newcommand{\de}{\mathrm{d}}
\newcommand{\vp}{\varphi}

\newcommand{\G}{{\cal G}}

\begin{document}

\title{Solar system constraints on $f(\G)$ gravity models}

\author{Antonio De Felice}
\affiliation{Theoretical and Mathematical Physics Group, Centre for
Particle Physics and Phenomenology, Louvain University, 2 Chemin du
Cyclotron, 1348 Louvain-la-Neuve, Belgium}

\author{Shinji Tsujikawa}
\affiliation{Department of Physics, Faculty
of Science, Tokyo University of Science, 1-3, Kagurazaka,
Shinjuku-ku, Tokyo 162-8601, Japan}

\date{\today}

\begin{abstract}
  We discuss solar system constraints on $f(\G)$ gravity models, where
  $f$ is a function of the Gauss-Bonnet term $\G$.  We focus on
  cosmologically viable $f(\G)$ models that can be responsible for
  late-time cosmic acceleration.  These models generally give rise to
  corrections of the form $\epsilon (r/r_s)^p$ to the vacuum
  Schwarzschild solution, where $\ve=H_*^2 r_s^2 \ll 1$, $r_s$ is the
  Schwarzschild radius of Sun, and $H_*$ is the Hubble parameter
  today.  We generally estimate the strength of modifications to
  General Relativity in order to confront models with a number of
  experiments such as the deflection of light and the perihelion
  shift.  We show that cosmologically viable $f(\G)$ models 
  can be consistent with solar system constraints for a wide range 
  of model parameters.
\end{abstract}

\maketitle

\section{Introduction}

The modified gravity paradigm has been extensively studied over the 
last few years as a way to make gravity responsible for the observed
acceleration of the universe at large scales \cite{review}. 
These models are attractive in the sense that cosmic acceleration 
can be realized without recourse to a dark energy matter component.
Unlike the cosmological constant scenario, they generally give rise to
a dynamical equation of state of dark energy that varies in time.

Among these modifications of gravity, the so called $f(R)$ theory
has become popular and started to be a new branch 
of cosmology \cite{Capo}.
In this theory the Lagrangian density $f$ is function of the Ricci scalar $R$. 
The $f(R)$ theory in the metric formalism can be proven to be classically 
equivalent to a class of Brans-Dicke theory with a Brans-Dicke parameter
$\omega_{\rm BD}=0$ \cite{Chiba}.
A scalar-field degree of freedom, called {\it scalaron} \cite{Star80}, 
can freely propagate to mediate a fifth force, 
unless the scalaron mass is heavy in the region of high density. 
For the compatibility with local gravity experiments 
the $f(R)$ theory needs to approach the Lagrangian density 
$f(R)=R-2\Lambda$ in General Relativity (GR) for large values of $R$
much larger than the present cosmological Ricci scalar $R_0$ \cite{lgcfR}.
Meanwhile non-trivial deviation from the $\Lambda$CDM model
can arise for $R$ of the order of $R_0$ \cite{AGPT,Li,AT08,Hu07,Star07,otherref}.
This leads to a number of interesting observational 
signatures such as the modifications to the matter 
power spectrum \cite{matterfR} as well as to the 
weak lensing spectrum \cite{lensingfR}.

It is also possible to take into account a Gauss-Bonnet (GB) scalar $\G$
that is a combination of the Ricci scalar $R$, the Ricci tensor 
$R_{\mu\nu}$, and the Riemann tensor $R_{\mu \nu \alpha \beta}$ \cite{GBghost,GenGh}.
This GB scalar, together with $R$, belongs to an infinite class of curvature invariants, 
the Lovelock scalars, which have the property that they do not introduce 
derivatives terms higher than two into the equations of motions for the metric tensor.
Among these scalars, $R$ and $\G$ are the only ones that do not identically 
vanish in four dimensions (apart from the cosmological constant). 
However the term $\sqrt{-g}\,\G$ is a total derivative in four dimensions, 
where $g$ is the determinant of the metric tensor.
This means that the only way for the GB term to contribute to the 
equations of motion is to couple it to something else, 
e.g.,\ a scalar field $\phi$ with the coupling of 
the form $F(\phi)\G$ \cite{GBpapers}.
This kind of coupling is present in the low energy effective 
action of string-theory \cite{Gasperini}, 
due to the presence of dilaton-graviton mixing terms.

The dynamics of dark energy based on the dilatonic coupling 
$F(\phi) \propto e^{\mu \phi}$ with the exponential potential 
$V(\phi) \propto e^{-\lambda \phi}$ has been studied by a number 
of authors \cite{NO05,Koi,TS,Sanyal,Neupane}. 
While it is possible to realize a scaling matter era \cite{Koi,TS} 
followed by a late-time cosmic acceleration, the region of 
viable parameter space to satisfy several observational constraints
is restricted to be very small \cite{Koi}. 
It was also shown in Refs.~\cite{TS,Ohta} that tensor perturbations tend to 
exhibit negative instabilities if the GB term is responsible for cosmic acceleration.
Moreover, in such models, the energy fraction of the GB term
needs to be strongly suppressed for the compatibility with local gravity 
experiments \cite{Amendavis}, which is at odds with the requirement 
of cosmic acceleration induced by the GB term.

There is another class of modified gravity models in which the Lagrangian 
density is described by $R+f(\G)$ (so-called ``$f(\G)$ gravity''), 
where $f$ is function in terms of the GB term $\G$ \cite{O05}. 
Unlike $f(R)$ gravity, this theory does not have an action 
in the Einstein frame with a standard kinetic term of 
a scalar-field degree of freedom. 
The conditions for cosmological viabilities of $f(\G)$
gravity have been studied in Refs.~\cite{LBM,CFG,ZCS}
(see also Refs.~\cite{Nojiri07,ULT}).
Li {\it et al.} \cite{LBM} showed that the condition $0<H^6f_{,\G\G} \ll 1$
(where $f_{,\G\G}\equiv \de^2f/\de \G^2$) needs to be fulfilled
in order to keep cosmological perturbations under control.
In Ref.~\cite{CFG} the same condition has been derived to ensure 
the stability of a late-time de Sitter solution as well as the existence 
of standard radiation and matter dominated epochs.
In particular the stability of the de Sitter point requires the condition 
$0<H_1^6 f_{,\G \G}<1/384$, where $H_1$ is the Hubble parameter
at this point. In order to remove possible singularities in
the cosmic expansion history the second derivative $f_{,\G \G}$
should not change sign, i.e.\ $f_{,\G\G}>0$ for all $\G$,  
together with the condition that $f_{,\G\G} \to 0$ as $|\G| \to\infty$.
This removes the presence of unstable modes during the whole past evolution
of the universe. A number of cosmologically viable models satisfying 
these requirements have been proposed in Ref.~\cite{CFG}.

In this paper we will study the property of $f(\G)$ gravity on solar 
system scales and investigate whether cosmologically viable $f(\G)$ 
models can be consistent with solar system constraints.
We first find approximate vacuum solutions for these
models through an iterative method introduced in Ref.~\cite{iterSol}. 
The solutions look like corrections to the \swa\ solution, where the
corrections are typically in the form of positive powers in the ratio
$r/r_s$, where $r$ is the distance from the center of compact object
and $r_s$ is the \swa\ radius. This behavior of positive powers is similar
to the typical correction that the cosmological constant 
gives to the \swa\ solution ($\propto (r/r_s)^2$). 
In the case of the \swa-de Sitter solution the metric in the solar system 
is dominated by the term $r_s/r$, but one can put upper bounds on the 
value of the cosmological constant such that its contribution is 
allowed by experimental data. We follow a similar procedure 
in order to constrain the values of model parameters 
in $f(\G)$ gravity.
We will show that cosmologically viable $f(\G)$ models can satisfy 
solar system constraints for a wide range of parameter space.

In section II we briefly review cosmologically viable $f(\G)$ models. 
In section III we describe the method used to find approximate
spherically symmetric solutions of the Einstein equations. 
In section IV we discuss solar system constraints in the presence of
positive powers of the radius correction to the \swa\ solution. 
In sections V, VI, and VII we apply the constraints to a number of
$f(\G)$ models. In section VIII we report our conclusions.

\section{Cosmologically viable $f({\cal G})$ models}

Let us first briefly review cosmologically viable $f({\cal G})$ models 
proposed in Ref.~\cite{CFG}.
The action describing this theory is given by 
\begin{equation}
\label{action}
S =  \int {\rm d}^4 x\sqrt{-g} 
\left[ \frac12\, R +f(\G) \right]
+S_m (g_{\mu \nu}, \Psi_m)\,,
\end{equation}
where $R$ is a Ricci scalar, $\G=R^2-4R_{\mu \nu}R^{\mu \nu}+
R_{\mu \nu \alpha \beta} R^{\mu \nu \alpha \beta}$ is 
a Gauss-Bonnet (GB) term, and $S_m$ is a matter action that depends
on a spacetime metric $g_{\mu \nu}$ and matter fields $\Psi_m$.
We use the unit $M_{\rm pl}=1/\sqrt{8 \pi G_N}=1$, 
but we restore the reduced Planck mass $M_{\rm pl}$ and 
the gravitational constant $G_N$ if required.

The late-time cosmic acceleration can be realized by the presence 
of a de Sitter (dS) point satisfying the condition 
$3H_1^2={\cal G}_1f_{,{\cal G}}({\cal G}_1)-f({\cal G}_1)$, where
$H_1$ and ${\cal G}_1$ are the Hubble parameter and the GB term 
at the dS point respectively.
The stability of the dS point demands
the condition $0<H_1^6 f_{,\G \G} (H_1)<1/384$ \cite{CFG}.
The GB term, $\G=24H^2(H^2+\dot{H})$, changes sign from negative to 
positive during the transition from the matter era to the accelerated epoch.
For the existence of standard radiation and matter eras we require that 
$f_{,\G \G}>0$ for $\G \le \G_1$ and that $f_{,\G \G}$ 
approaches $+0$ in the limit $|\G| \to \infty$.
We also need the regularities of $f(\G)$ and its derivatives $f_{,\G}$, $f_{, \G\G}$.
The following two models can satisfy these conditions \cite{CFG}:
\begin{eqnarray}
& &{\rm (A)}~f (\G)=\lambda \frac{\G}{\sqrt{\G_*}}\,{\rm arctan} \left( \frac{\G}{\G_*} \right)
-\frac{1}{2}\lambda \sqrt{\G_*}\,{\rm ln} \left(1+\frac{\G^2}{\G_*^2} \right)
-\alpha \lambda \sqrt{\G_*} \label{modela} \,,\\
& &{\rm (B)}~f (\G)=\lambda \frac{\G}{\sqrt{\G_*}}\,{\rm arctan} \left( \frac{\G}{\G_*} \right)
-\alpha \lambda \sqrt{\G_*} \label{modelb}\,,
\end{eqnarray}
where $\alpha$, $\lambda$ and ${\cal G}_*$ are positive constants.
Note that ${\cal G}_*$ roughly corresponds to the scale $H_*^4$ 
for $\alpha$ and $\lambda$ of the order of unity, 
where $H_*$ is the Hubble parameter today.
The second derivative of $f$ with respect to $\G$ is 
$f_{,\G\G}=\lambda/[\G_*^{3/2} (1+\G^2/\G_*^2)]$ and 
$f_{,\G\G}=2\lambda/[\G_*^{3/2} (1+\G^2/\G_*^2)^2]$
for the models (A) and (B) respectively, 
so that $f_{,\G \G}>0$ for $\lambda>0$.

In the region of high density where local gravity experiments are 
carried out ($|\G| \gg \G_*$), 
the above models have the following asymptotic behavior
\begin{eqnarray}
& &{\rm (A)}~f (\G) \simeq \frac12 \pi \lambda \frac{\G}{\sqrt{\G_*}}
-(\alpha+1)\lambda \sqrt{\G_*}-\frac12\lambda \sqrt{\G_*}\,\ln 
\left( \frac{\G^2}{\G_*^2} \right)
-\frac{\lambda \sqrt{\G_*}}{6} \frac{\G_*^2}{\G^2}\,,
\label{asmodela}
\\
& &{\rm (B)}~f (\G)\simeq \frac12 \pi \lambda \frac{\G}{\sqrt{\G_*}}
-(\alpha+1)\lambda \sqrt{\G_*}+\frac{\lambda \sqrt{\G_*}}{3}
\frac{\G_*^2}{\G^2}
\label{asmodelb}\,.
\end{eqnarray}
The first terms in Eqs.~(\ref{asmodela}) and (\ref{asmodelb}) are linear in ${\cal G}$
so that they do not give rise to any contribution to the Einstein equation, whereas 
the second terms contribute to the field equation as a cosmological constant.
The other terms in Eqs.~(\ref{asmodela}) and (\ref{asmodelb}) correspond to the 
corrections to the $\Lambda$CDM model.
The difference between the models (A) and (B) is that the former has a logarithmic 
correction that mildly increases with the growth of $|\G|$.
Note that the viable $f(R)$ models such as 
(i) $f(R)=R-\lambda R_* (R/R_*)^{2n}/[ (R/R_*)^{2n}+1]$ and 
(ii) $f(R)=R-\lambda R_* [1-(1+R^2/R_*^2)^{-n}]$ ($n>0$) 
behave as $f(R) \simeq R-\lambda R_*+\lambda R_*(R_*/R)^{2n}$
in the region of high density ($R \gg R_*$). This asymptotic 
form is similar to the model (B) given above by replacing $R$ for $\G$. 

In the following we shall study solar-system constraints on cosmologically 
viable $f(\G)$ models. Before doing so, it is worth mentioning the difference 
between $f(R)$ and $f(\G)$ theories.
If we consider a spherically symmetric background, the \swa\ vacuum solution 
corresponds the vanishing Ricci scalar ($R=0$).
In the presence of non-relativistic matter, $R$ approximately 
equals to the matter density $\rho_m/M_{\rm pl}^2$ 
for viable $f(R)$ models \cite{Hu07,Star07,otherref}.
Then the term $(R_*/R)^{2n}$ is roughly of the order of $(\rho_c/\rho_m)^{2n}$, 
where $\rho_c$ is the cosmological density today.
The ratio $(\rho_c/\rho_m)^{2n}$ becomes much smaller than 1 for $n>0$
in the region of high density ($\rho_m \gg \rho_c$) so that one has 
$(R_*/R)^{2n} \ll 1$. In the presence of non-relativistic matter the chameleon 
mechanism \cite{chameleon} can be at work for the $f(R)$ models 
that have the asymptotic form $f(R) \simeq R-\lambda R_*+\lambda R_*(R_*/R)^{2n}$
in the region $R \gg R_*$, which allows the possibility for the consistency 
with local gravity tests. In fact it was shown in Ref.~\cite{CapoTsuji} that 
these models can satisfy  
solar system and equivalence principle constraints for $n > 0.9$.

On the contrary one has $\G=R_{\mu\nu\alpha\beta}R^{\mu\nu\alpha\beta}=
12\,r_s^2/r^6$ on the vacuum \swa\ solution, where $r_s=2G_NM_{\odot}$
is the \swa\ radius and $M_{\odot}$ is the mass of the star.
Since $\G$ does not vanish even in the vacuum, the term such as
$(\G_*^2/\G^2)^n$ ($n>0$) can be much smaller than 1 
even in the absence of non-relativistic matter.
If matter is present, this will give rise to the contribution of 
the order of $R^2 \approx (8\pi G_N \rho_m)^2$ to the 
GB term. The ratio of the matter contribution to 
the vacuum GB value $\G^{(0)}=12\,r_s^2/r^6=48(G_N M_{\odot})^2/r^6$
can be estimated as 
\begin{equation}
s \equiv \frac{R^2}{\G^{(0)}}  \approx \frac{(8\pi)^2}{48} 
\frac{\rho_m^2 r^6}{M_{\odot}^2}\,.
\end{equation}
As long as $s \ll 1$, we can neglect the matter contribution 
to the GB term.

At the surface of Sun (radius $r_{\odot}=6.96 \times 10^{10}$\,cm$=3.53\times 10^{24}$\,GeV$^{-1}$
and mass $M_{\odot}=1.99 \times 10^{33}\,{\rm g}=1.12 \times 10^{57}$\,GeV), 
the density $\rho_m$ drops down rapidly from the order $\rho_m \approx 10^{-2}$\,g/cm$^3$
to the order $\rho_m \approx 10^{-16}$\,g/cm$^3$.
If we take the value $\rho_m = 10^{-2}$\,g/cm$^3$ we have 
$s \approx 4 \times 10^{-5}$ (where we have used $1$\,g/cm$^3=4.31 \times 10^{-18}$\,GeV$^4$).
Taking the value $\rho_m = 10^{-16}$\,g/cm$^3$ leads to 
a much smaller ratio: $s \approx 4 \times 10^{-33}$.
The matter density approaches a constant value $\rho_m \approx 10^{-24}$\,g/cm$^3$
around the distance $r=10^3r_{\odot}$ from the center of Sun.
Even at this distance we have $s \approx 4 \times 10^{-31}$, 
which means that the matter contribution to the GB term can 
be completely neglected in the solar system we are interested in.
If we use the value $\rho_m \approx 10^{-24}$\,g/cm$^3$, 
$s$ exceeds the order of 1 for the distance $r \gtrsim 10^8 r_{\odot}$.
However this is out of the region where solar system experiments are concerned.
Moreover the \swa\ solution has no meaning far away from the
star where other contributions can arise, i.e.\ other close stars,
the mean field of the galaxy, and so on. 
{}From the above discussion we find that vacuum solutions can be used 
when we discuss solar system constraints on $f(\G)$ gravity.

\section{Expansion around the spherically symmetric spacetime}

The cosmologically viable $f(\G)$ models (\ref{modela}) and (\ref{modelb}) 
will consist of a numerical factor of order $\sqrt{\G_*}$ times
a dimensionless function (because $f(\G)$ has the dimension of [mass]$^2$).
Meanwhile the typical value of the GB term on the vacuum Schwarzschild solution 
is given by 
\begin{equation}
\G_s \equiv 12/r_s^4\,.
\end{equation}
When we discuss solar system constraints, it is convenient to define the 
following dimensionless ratio
\begin{equation}
\ve \equiv \sqrt{\frac{\G_*}{\G_s}}\,.
\label{vedef}
\end{equation}
Since $\sqrt{\G_*}$ is of the order of the squared of the present Hubble 
parameter $H_*$, the parameter $\ve$ is approximately given by 
$\ve \approx (H_* r_s)^2/(2\sqrt{3})$.
In the subsequent sections we shall discuss the case of Sun with 
the Schwarzschild radius $r_s=2.95 \times 10^3$\,m.
Using the value $H_* \approx 70$\,km\,sec$^{-1}$\,Mpc$^{-1}$, 
the parameter $\ve$ for Sun is approximately given by 
\begin{equation}
\varepsilon \approx 10^{- 46}\,.
\label{epvalue}
\end{equation}
The model (\ref{modela}) can be written in the form
$f(\G)=\ve \left[ \lambda \sqrt{\G_s}g(x)-\alpha \lambda \sqrt{\G_s} \right]$, 
where $g(x)=x \arctan x-(1/2)\ln (1+x^2)$ and 
$x=\G/\G_*=\G/(\G_s \epsilon^2)$.
Hence the function $f$ can be replaced by the form $f=\ve \tilde{f}$.

The equations of motion in the vacuum can be written as
\begin{equation}
G^\mu{}_{\nu}+\ve\,\Sigma^\mu{}_{\nu}=0\,,
\label{geeq}
\end{equation}
where $G^\mu{}_{\nu}$ is the Einstein tensor, and 
\begin{equation}
\label{eq:VC2}
\Sigma_{\mu\nu}=  8 \left[ R_{\mu \rho \nu \sigma} +R_{\rho \nu} g_{\sigma \mu}
-R_{\rho \sigma} g_{\nu \mu} -R_{\mu \nu} g_{\sigma \rho}+
R_{\mu \sigma} g_{\nu \rho}+R(g_{\mu \nu} g_{\sigma \rho}
-g_{\mu \sigma} g_{\nu \rho})/2 \right] \nabla^{\rho} \nabla^{\sigma} \tilde{f}_{,\G}
+(\G \tilde{f}_{,\G}-\tilde{f}) g_{\mu \nu}.
\end{equation}
In general these theories will have vacuum solutions, 
which we wish to study here. 
Although exact solutions are not always found analytically,
it is possible to obtain approximate solutions which reproduce
the real ones very well at least on some scales. 
In fact, since $\ve\ll1$, we can try to use the iterative method 
introduced in Ref.~\cite{iterSol}. 

We look for static spherical symmetric solutions of the kind
\begin{equation}
\de s^2=-A(r,\ve)\,\de t^2+\frac{\de r^2}{B(r,\ve)}
+r^2\,(\de\theta^2+\sin^2\theta\,\de\varphi^2)\,,
\label{eq:genmet}
\end{equation}
where the functions $A$ and $B$ are Taylor expanded in the form:
\begin{align}
\label{eq:VCB}
A(r,\ve) &=A_0(r)+A_1(r) \ve+A_2(r)\,\ve^2+\dots\,, \\
B(r,\ve) &=B_0(r)+B_1(r) \ve+B_2(r)\,\ve^2+\dots\,.
\end{align}
Using this expansion for $A$ and $B$, both
$G^\mu{}_{\nu}(r,\ve)$ and $\Sigma^\mu{}_{\nu}(r,\ve)$ 
can be expanded order by order in $\ve$. 
At lowest level, the equations of motion correspond to
\begin{equation}
\label{eq:VC4}
G^\mu{}_\nu{}^{(0)}=0\, ,
\end{equation}
which leads to the usual \swa\ solution $B_0=A_0=1-r_s/r$.
At first order one has
\begin{equation}
\label{eq:VC5}
\ve\,[G^\mu{}_\nu{}^{(1)}(A_1,B_1,A_0,B_0)+
\Sigma^\mu{}_\nu{}^{(0)}(A_0,B_0)]=0\,,
\end{equation}
which reduces to ordinary differential equations for $A_1,B_1$, where $A_0,B_0$ are
the \swa\ terms found previously. This method can be reiterated to
get the coefficients $A_2,B_2$, as well as all the other 
higher-order terms.

It should be noted that for general approach to the modification of
gravity, including quintessence, there is no more analogue of the
Birkhoff theorem regarding the unicity of the solution for a
spherically symmetric vacuum solution. 
We can only impose to have a static metric.
The bottom line is that the forms of $A$ and $B$ can be found at any
order-$\ve$ approximation, by solving the Einstein equations
iteratively with respect to the small parameter $0<\ve \ll 1$.

Suppose that we find such an iterative solution and write both $A$ 
and $B$ as power expansions of $\ve$. 
Then the iterative solution should have in general a radius of convergence, 
inside which each correction is larger than the next order $\ve$ term. 
In this case we expect that the dominant term corresponds to the \swa\ contribution. 
Therefore in the small $\ve$ limit we will have
\begin{align}
A&=1-1/\rho+\ve\,c_1\,\rho^p+\ve^2\,c_3\,\rho^m+{\cal O}(\ve^3)\,,
\label{eq:pertA1}\\
B&=1-1/\rho+\ve\,c_2\,\rho^q+\ve^2\,c_4\,\rho^n+{\cal O}(\ve^3)\,,
\label{eq:pertA2}
\end{align}
where $\rho\equiv r/r_s$. 
In Secs.~\ref{modelasec} and \ref{modelbsec} we will show that the correction 
terms in Eqs.~(\ref{eq:pertA1}) and (\ref{eq:pertA2}) in fact arises for 
the cosmologically viable $f(\G)$ models given in (\ref{modela}) and (\ref{modelb}).
We will restrict ourselves to the region $\rho\gg1$, which is 
generically satisfied in the solar system outside Sun.
In other words this corresponds to the weak limit of the theory.
We have introduced $c_{1,2,3,4}$ as constants whose values need to be bounded
experimentally, and also $p,q,m,n$ as the powers of $\rho$. 
In the following we will restrict our attention to the case $p=q>0$ and $m=n>0$.
Later we will see that this assumption is quite
general for cosmologically viable $f({\cal G})$ models. 

In order to have a meaningful $\ve$ expansion, one needs to verify that 
there exists a set of convergence, e.g., 
\begin{equation}
\label{eq:espan}
D=\{\rho\ |\ \rho\gg1,\,\ve\rho^p\ll\rho^{-1} \}\,.
\end{equation}
This implies that the expansions (\ref{eq:pertA1}) and (\ref{eq:pertA2}) 
can be trusted only in this region. Outside it, one must find the solutions of Einstein equations, 
both analytically or numerically, without using the $\ve$ expansion.
This also implies that it does not make sense to study this solution in the limit
$\rho\to\infty$. However, even if the full solution is known, it
does not have physical meaning for values of $r$ where the solar
system begins to feel other stars' contribution. This situation is
analogous to the \swa-de Sitter solution, where the same solution cannot
be trusted any more at distances a few parsecs away from Sun.
In this region the solar system cannot be treated as an entity 
isolated from the rest of the galaxy. 
Of course, for larger $p$, the set $D$ becomes smaller.
However, this is not enough: one should also verify that in the same set the
$\ve^2$-term is much smaller than the $\ve$-one. 
This implies that in $D$ one also requires to have
\begin{equation}
 D_{2}=\{\rho\ |\ \rho\gg1, \ve\rho^m\ll\rho^p \}\subseteq D\, .
\end{equation}
If $0\leq m\leq 2p+1$ 
then this condition is verified. We will call a
good $\ve$ expansion for the one where this last condition is
valid. In general we may have more complicated cases where there 
exists an order of expansion $d$ such that $D_i\subseteq D_d$ for all
$i\geq1$ (and $d$ might not be one), such that the expansion makes sense in $D_d$. If there is no
such set, then the expansion does not make sense. If there is such a set, 
we can define $D=D_d$.

If $D_2\subseteq D$, then, in $D$ (or, more in general, in $D_d$), it
is safe to approximate the perturbative solution as
\begin{equation}
A=1-1/\rho+\ve\,c_1\,\rho^p\,,\qquad
B=1-1/\rho+\ve\,c_2\,\rho^p\,.
\label{REDG}
\end{equation}
In the next section we shall study a number of solar system bounds
for the metric (\ref{REDG}).

\section{Solar system constraints}
\label{solarconst}

There are a number of solar system constraints on the deviation from 
General Relativity--such as (a) deflection of light, (b) Cassini experiment, 
(c) Perihelion shift, (d) retardation of light, and (e) gravitational redshift.
In the following we discuss those constraints for a general metric 
in the form of (\ref{REDG}).

\subsection{Deflection of light}
\label{deflection}

The first constraint we discuss is the deflection of light. 
The Lagrangian for a photon moving in the $\theta=\pi/2$ plane
in the gravitational field of the metric (\ref{eq:genmet}) is given by
\begin{equation}
L=\tfrac12 A \dot t^2-\tfrac12 B^{-1}\dot r^2-\tfrac12 r^2 \dot\vp^2=0\,,
\end{equation}
where a dot represents an affine parameter along the geodesics. 
There are two constants of motion, namely
\begin{equation}
E\equiv A\dot t\qquad{\rm and}\qquad
L\equiv r^2\dot\vp\,.
\label{ELdef}
\end{equation}
Then we find 
\begin{equation}
\dot r^2=L^2\left(\frac{E^2}{L^2}\frac BA-\frac B{r^2}\right)\,.  
\label{eq:luce1}
\end{equation}
The minimal distance $r_0$ can be defined 
such that $\dot r(r_0)=0$, giving
\begin{equation}
\frac{L^2}{r_s^2 E^2}=\frac{\rho_0^2}{A(\rho_0)}\,,
\label{Lsre}
\end{equation}
where $\rho_0 \equiv r_0/r_s$.

Integrating $\de\vp/\de r=\dot\vp/\dot{r}$
by using Eqs.~(\ref{ELdef})-(\ref{Lsre}), we obtain
\begin{equation}
\vp(\rho)=\pm\int_{\rho_0}^\rho \de\bar\rho\,\frac{\rho_0}
{\bar\rho}\,\sqrt{\frac{A(\bar\rho)}
{B(\bar\rho)[A(\rho_0)\bar\rho^2-A(\bar\rho)\rho_0^2]}}\,,
\label{eq:phia}
\end{equation}
where $\bar{\rho}$ is used to avoid the confusion with the upper 
limit $\rho$ of the integral.
The Schwarzschild-de Sitter solution corresponds to $c_1=c_2$ and $p=2$,
in which case the integral (\ref{eq:phia}) reduces to the standard GR contribution.
Therefore the cosmological constant does not 
give any modification to GR for light bending \cite{NoLight}. 
In standard GR, the integral is taken in the limit $\rho\to\infty$. 
This is a mathematical extrapolation, as the metric does not hold in the whole spacetime. 
Of course this property holds in our approach where the approximate metric is sensible 
only in $D$, the domain of convergence. 
Up to first order in $\ve$ and in the domain $D$, 
it is possible to approximate the integral as
\begin{equation}
\vp=\pm\int_{\rho_0}^\rho \de\bar\rho\,\frac{\rho_0}{\bar\rho}
\sqrt{\frac{\bar\rho\,\rho_0}{\bar\rho^3(\rho_0-1)-\rho_0^3(\bar\rho-1)}}
+\ve\int_{\rho_0}^\rho \de\bar\rho\,F(\rho_0,\bar\rho)\,,
\end{equation}
where 
\begin{equation}
\label{eq:Flt}
F(\rho_0, \rho)=\left[\frac{\rho  \rho_0}{\rho ^3
(\rho_0-1)-\rho  \rho_0^3+\rho_0^3}\right]^{3/2} \frac{\rho ^p \bigl\{c_1 \rho ^3
(\rho_0-1)+c_2 \bigl[\rho ^3 (1-\rho_0)+(\rho -1) \rho_0^3\bigr]\bigr\}
-c_1 (\rho -1) \rho ^2 \rho_0^{p+1}}{2\rho(\rho-1)}\, .
\end{equation}
We can further simplify this expression 
by considering the limits $\rho,\rho_0\gg1$:
\begin{equation}
\vp=\pm\int_{\rho_0}^\rho \de\bar\rho\,\frac{\rho_0}{\bar\rho\sqrt{\bar\rho^2-\rho_0^2}}
\pm\int_{\rho_0}^\rho \de\bar\rho\,
\frac{\bar\rho^2+\rho_0\bar\rho+\rho_0^2}{2\bar\rho^2(\bar\rho+\rho_0)
\sqrt{\bar\rho^2-\rho_0^2}}\pm \frac{1}{2} \ve\rho_0\int_{\rho_0}^\rho \de\bar\rho\,
\frac{c_1\bar\rho^2(\bar\rho^p-\rho_0^p)-c_2\bar\rho^p(\bar\rho^2-\rho_0^2)}
{\bar\rho(\bar\rho^2-\rho_0^2)^{3/2}}\,.
\label{vpint}
\end{equation}
Taking the positive sign in front of each integral, the deviation
angle in the region $D$ is given by
\begin{equation}
\vartheta(\rho)=2\int_{\rho_0}^\rho \de\bar\rho\,\frac{\rho_0}{\bar\rho\sqrt{\bar\rho^2-\rho_0^2}}
+\int_{\rho_0}^\rho \de\bar\rho\,
\frac{\bar\rho^2+\rho_0\bar\rho+\rho_0^2}{\bar\rho^2(\bar\rho+\rho_0)\sqrt{\bar\rho^2-\rho_0^2}}
+\ve\,\rho_0\int_{\rho_0}^\rho \de\bar\rho\,
\frac{c_1\bar\rho^2(\bar\rho^p-\rho_0^p)-c_2\bar\rho^p(\bar\rho^2-\rho_0^2)}
{\bar\rho\,(\bar\rho^2-\rho_0^2)^{3/2}}-\pi\, ,
\label{eqzLD}
\end{equation}
in the limit that $\rho\gg\rho_0$. 
The first two contributions in Eq.~(\ref{vpint}), 
corresponding to the GR ones, give
\begin{align}
\vartheta_\mathrm{GR}=\frac{2}{\rho_0}+{\cal O}(\rho_0/\rho)\,.
\label{vartheGR}
\end{align}
Meanwhile the $\ve$-contributions can be evaluated as
\begin{eqnarray}
& &  p=1,\qquad \vartheta_\ve=-\ve\,c_1\,\rho_0+
\ve\rho_0(c_1-c_2)\ln(2\rho/\rho_0)+{\cal O}(\rho_0^{-1})\,,\label{peq1}\\
&&   p=2,\qquad \vartheta_\ve=(c_1-c_2)\ve\,\rho_0\rho+{\cal O}(\rho_0^{-1})\,,\\
& &  p=3,\qquad \vartheta_\ve=\tfrac12(c_1-c_2)\ve\,\rho_0\,\rho^2
+\bar c\,\ve\,\rho_0^3\,[\ln(2\rho/\rho_0)-1]+{\cal O}(\rho_0^{-1}),\label{peq3}\\
& & p \geq 4,\qquad \vartheta_\ve
=\frac{c_1-c_2}{p-1}\,\ve\,\rho_0\,\rho^{p-1}+\frac{\bar c}{p-3}\,\ve\,\rho_0^3\,
\rho^{p-3}+{\cal O}(\rho_0^{p-5})\,.\label{peq4}
\end{eqnarray}
We have kept not only the dominant correction but also the next one. 
Of course the latter becomes important only for $c_1=c_2$, that is,
when the first correction vanishes. 
Therefore we have evaluated the second (smaller) contribution only when the first vanishes, 
that is, when $c_1=c_2$. In other words, we define $\bar c=c_1$ if $c_1=c_2$. 
The experimental bound on $\vartheta_\ve$ relative to $\vartheta_\mathrm{GR}$
is given by the Very Long Baseline Interferometry (VLBI), which combined measurements 
taken by different radio telescopes on Earth \cite{VLBI}. 
The experimental value $\vartheta_{\rm exp}$ relative to the 
theoretical prediction has been constrained to be 
$\vartheta_{\rm exp}/\vartheta_{\rm theor}=1.0001 \pm 0.0001$. 
Therefore, in order that $\vartheta_\ve$ does not affect the GR result,
we impose
\begin{equation}
\frac{\vartheta_\ve}{\vartheta_\mathrm{GR}}<10^{-4}\,.
\label{deflectionbound}
\end{equation}
When $p \ge 2$ this condition translates into
\begin{equation}
\frac{c_1-c_2}{2(p-1)}\ve\rho_0^2\rho^{p-1}<10^{-4}\,.
\label{eq:Lct}
\end{equation}
Recall that in the domain $D$ we have
\begin{equation}
\label{eq:dmnL}
\ve\,\rho_0\rho^{p-1}\ll\frac1{\rho_0}\,
\frac{\rho_0^2}{\rho^2}\ll\frac1{\rho_0}\,,
\end{equation}
in which case the condition (\ref{eq:Lct}) is satisfied for $c_1$, $c_2$, $p$
of the order of unity.
Hence the deflection of light always remains a small correction
in the domain of convergence.

\subsection{Cassini experiment}

Iess {\it et al.} \cite{Bertotti} showed that the contribution to the fractional frequency shift of a stable and coherent two-way radio signal (Earth-spacecraft-Earth) $y$, due to the metric of a gravitational theory (which possesses a weak field limit),
is proportional to the deviation angle $\vartheta$ of light, and it is given by the formula
\begin{equation}
\label{eq:Cs1}
y=2\,\frac{v_1l_0+v_0l_1}{l_0+l_1}\,\vartheta \,,
\end{equation}
where $v_0$ and $v_1$ are transverse velocities of Earth and a spacecraft, 
and $l_0$ and $l_1$ are their distances from Sun. 
Since $l_1\gg l_0$, the $\ve$ contribution can be written as
\begin{align}
y_\ve&\approx 2\,v_{\rm Earth}\,[\vp(\rho_0,\rho_{\rm Earth})
+\vp(\rho_0,\rho_{\rm Cassini})]\notag\\
&\approx v_{\rm Earth}\,\frac{c_1-c_2}{p-1}\,\ve\,
\rho_0\,(\rho^{p-1}_{\rm Cassini}
+\rho^{p-1}_{\rm Earth})\,,
\label{eq:Cs2}
\end{align}
where the approximate equality in the second line is valid for $p \ge 2$.
Meanwhile, from Eq.~(\ref{vartheGR}), the GR contribution can be written as
\begin{equation}
\label{eq:Cs3}
y_{\rm GR} \approx 4\,\frac{v_{\rm Earth}}{\rho_0}\, .
\end{equation}
The $\ve$ contribution needs to be a negligible correction to GR.
In order to have $y_\ve \ll y_{\rm GR}$ we require that
\begin{equation}
\ve \rho_0^2 \rho_{\rm Cassini}^{p-1} \ll 1\,,
\end{equation}
which is satisfied, as we have seen before, in the region $D$, 
of course if $\rho_{\rm Cassini}\in D$. 

The signal due to GR detected by Cassini is $y\sim10^{-10}$, within an experimental error 
of $\Delta y_{\rm exp}\sim10^{-14}$. Therefore the contribution $y_\ve$ from the 
modifications of gravity needs to satisfy the condition
$y_{\ve}<\Delta y_{\rm exp}\sim 10^{-14}$, 
or equivalently, $y_\ve/y_{\rm GR}<10^{-4}$. 
If $p \ge 2$, this condition translates into 
\begin{equation}
\frac{c_1-c_2}{4(p-1)} \ve \rho_0^2
\rho_{\rm Cassini}^{p-1}<10^{-4}\,.
\label{cassibo}
\end{equation}
As we have shown in Eq.\ (\ref{eq:dmnL}), this relation is satisfied 
in the domain of convergence. 
The constraint (\ref{cassibo}) can be used to place 
experimental bounds on $f(\G)$ models later.

\subsection{Perihelion shift}

Let us proceed to constraints coming from the perihelion shift of the inner planets, 
whose GR value is in extremely good agreement with experimental data. 
We will follow a similar procedure to the one discussed in subsection \ref{deflection}. 
The difference is that we now deal with 
the motion of a massive particle whose Lagrangian reads
\begin{equation}
\label{eq:lagMass}
L=\tfrac12A\dot t^2-\tfrac12\,B^{-1}\,\dot r^2-\tfrac12\,r^2\dot\vp^2=\tfrac12 .
\end{equation}
As in the previous case, the constants of motion are
\begin{equation}
E=A\dot t\qquad\textrm{and}\qquad
L=r^2 \dot\vp\,.
\end{equation}
Defining $u=r_s/r$, we can derive the following differential equation
\begin{equation}
\frac{\de^2u}{\de\vp^2}+u-\frac{r_s^2}{2L^2}
=\frac{r_s^2E^2}{2L^2}\frac BA\left( \frac1B\frac{\de B}{\de u}
-\frac1A\frac{\de A}{\de u}\right)
-\frac12\frac{\de B}{\de u}u^2-(B-1)u-\frac{r_s^2}{2L^2}
\left( \frac{\de B}{\de u}+1\right)\,.
\label{eq:vid1}
\end{equation}

The flat-space solution can be obtained by setting the right hand side of Eq.~(\ref{eq:vid1}) to be zero, as
\begin{equation}
\label{eq:vid}
u^\flat=\frac{r_s^2}{2L^2}\,[1+\delta\cos(\phi-\phi_0)]\,,
\end{equation}
where $0<\delta<1$ is the eccentricity of the closed orbit. 

From Eq.~(\ref{eq:lagMass}) it follows that 
\begin{equation}
\dot{r}^2=B \left( \frac{E^2}{A}-\frac{L^2}{r^2}-1 \right)\,.
\end{equation}
The minimum distance $\rho_0=r_0/r_s$ satisfies
\begin{align}
\frac{r_s^2 E^2}{2L^2} &= A(\rho_0) \left( \frac{1}{2\rho_0^2}
+\frac{r_s^2}{2L^2} \right) 
\nonumber \\
&= \left( 1-\frac{1}{\rho_0}+\ve\, c_1 \rho_0^p \right)
\left( \frac{1}{2\rho_0^2}+\frac{r_s^2}{2L^2} \right) 
\simeq \frac{1}{2\rho_0^2}
+\frac{r_s^2}{2L^2}\,.
\label{rse}
\end{align}
Since both $E$ and $L$ are constants, Eq.~(\ref{rse}) is an algebraic condition 
for $\rho_0$. Therefore, once
we fix the orbit, that is $r_s$, $E$ and $L$, then we also fix
$r_0$. The same solution will be valid at all times, therefore the
orbit will always have the same condition for $r_0$, i.e.\ the same
perihelion and the same aphelion. 
In other words, the minimum (and the maximum as well) value 
for $r$ will be unchanged at successive perihelia. 
Since $\dot r=0$ also implies $\de r/\de\vp=0$, the same
initial conditions for the differential equation at the perihelia are identical 
at each perihelion and so the orbit repeats exactly, 
see Rindler's book \cite{Rindler}.

Expanding Eq.~(\ref{eq:vid1}) at linear order in $\ve$, we find
\begin{equation}
\label{eq:vid1b}
 \frac{\de^2u}{\de\vp^2}+u-\frac{r_s^2}{2L^2}=\frac32 u^2+
\frac{\ve  u^{-p-1}r_s^2}{2 L^2 \rho_0^3} \left\{
c_1 p (\rho_0-1) \left(L^2r_s^{-2}+\rho_0^2\right)+c_2
\{L^2r_s^{-2} \left[(p-2) u^2 \rho_0^3-p (\rho_0-1)\right]+p
\rho_0^2\}\right\}\,.
\end{equation}
We evaluate the right hand side of Eq.~(\ref{eq:vid1b}) on 
the Newtonian solution $u^\flat$. 
In this case, using Eq.~(\ref{eq:vid}), we have
\begin{equation}
\frac{L^2}{r_s^2}\approx \frac12\,\rho_0\,(1+\delta)\,.
\label{eq:rho0}
\end{equation}
By doing so, Eq.~(\ref{eq:vid1b}) reduces to 
\begin{align}
  \frac{\de^2u}{\de\vp^2}+u-\frac{r_s^2}{2L^2}&=
\frac{\ve\,[\rho_0(1+\delta)]^p}
{2 (\delta +1) \rho_0^2 (\delta  \cos\vp+1)^{p+1}} 
\bigl\{c_1p (\delta +1) (\rho_0-1)(\delta +2 \rho_0+1)+
c_2(p-2)\delta ^2 \rho_0 \cos ^2\vp\notag\\
&~~~+c_2 \{p [\delta ^2 (1-\rho_0)+2 \delta +2 \rho_0+1]-2 \rho_0\}
+2 c_2 (p-2) \delta  \rho_0 \cos\vp\bigr\}\, .
\end{align}
In the limit $\rho_0\gg1$, the right hand side can be simplified to
\begin{align}
\frac{\de^2u}{\de\vp^2}+u-\frac{r_s^2}{2L^2}&=
\frac{\ve\,[\rho_0(1+\delta)]^p}
{2 (\delta +1) \rho_0^2 (\delta  \cos\vp+1)^{p+1}}
\{2 c_1 p\,(1+\delta)\rho_0^2\notag\\
&~~~+\rho_0 [c_1 p\, (\delta^2-1)+c_2 (p-2) \delta ^2 \cos ^2\vp
-c_2 \bigl(p (\delta ^2-2)+2\bigr)+2
c_2 (p-2) \delta  \cos \vp]\}\,.
\label{eq:appPE}
\end{align}
The second term on the right hand side of Eq.~(\ref{eq:appPE}) is subdominant unless $c_1=0$. 
In the following we will focus on the case $c_1\neq0$.

The solutions can be written down and studied for each $p$. 
As it happens in GR, the homogeneous solution can be described by
the Newtonian solution with some periodic corrections, 
i.e.\ only dependent on $\cos\vp$ and higher harmonics ($\cos2\vp$ and so on). 
However there will be terms which have a secular impact on the orbit, 
and here we are looking exactly for such terms.
The solution for Eq.~(\ref{eq:appPE}), valid for any $p$,
is given by 
\begin{align}
u=\frac{r_s^2}{2L^2}\,[1+\delta\,\cos\vp]
+\frac{3\delta\,\vp\sin\vp}{2(1+\delta)^2\rho_0^2}
-\delta\,\ve\,p(p+1)\,\frac{c_1\,\rho_0^p\,P_1(\delta^2)}
{\sqrt{1-\delta^2}\,(1-\delta)^p}\sin\vp\,\arctan\!
\left[\sqrt{\frac{1-\delta}{1+\delta}}\tan\frac\vp2\right]
+P_2(\delta,\vp)\,,
\label{eq:szp}
\end{align}
where $P_1(x)$ is a polynomial such that $P_1(0)=1$ with degree in $x$ equal to 
the integer part of $p/2$, and $P_2$ is periodic in $\vp$ so that it does not have 
any secular contribution. The first contribution on the right hand side of Eq.~(\ref{eq:szp})
is the Newtonian solution, the second one is the standard GR secular correction, 
whereas the third one is the secular contribution due to the $\ve$-modification of the metric. 
Let us examine orbits with small eccentricity. 
Then at lowest order in $\delta$ one finds 
that Eq.~(\ref{eq:szp}) reduces to
\begin{equation}
u\approx\frac{r_s^2}{2L^2}\,[1+\delta\,\cos\vp]
+\frac{3 \delta\,\vp\sin\vp}{2(1+\delta)^2\rho_0^2}
-\frac12\,\delta\,\ve\,c_1\,p\,(p+1)\,\rho_0^p\,\vp\,\sin\vp\,,
\label{uapso}
\end{equation}
where we have neglected a possible constant contribution from the arctan 
which, multiplied by $\sin\vp$, reduces to a periodic term. 

The result (\ref{uapso}) coincides with another simpler method 
at lowest order in $\delta$. 
One can expand Eq.~(\ref{eq:appPE}) in series of $\delta$, giving
\begin{equation}
\label{eq:deltPER}
\frac{\de^2u}{\de\vp^2}+u-\frac{r_s^2}{2L^2}=
c_1\, p\, \ve\, \rho_0^p\, [1-(p+1) \delta  \cos\vp+p\,\delta]
+\frac{6\, \delta\,  (\cos\vp-1)+3}{2 \rho_0^2}+{\cal O}(\delta^2)\,.
\end{equation}
The solution of this equation is 
\begin{equation}
u\approx\frac{r_s^2}{2L^2}\,[1+\delta\,\cos\vp]
+\frac{3\delta\,\vp\sin\vp}{2\rho_0^2}
-\frac12\,\delta\,\ve\,c_1\,p\,(p+1)\,\rho_0^p\,\vp\,\sin\vp
+d_1\cos(2\vp)+d_2\cos(3\vp)\,,
\label{eq:deltPER2}
\end{equation}
where the coefficients $d_{1,2}$ are not needed to be known for our
purpose.  The second term of Eq.~(\ref{eq:deltPER2}) coincides with
the second term of Eq.~(\ref{uapso}) at lowest order in $\delta$. The
corrections are due to higher orders of $\delta$, 
which are not included in Eq.~(\ref{eq:deltPER2}).

Using the relation (\ref{eq:rho0}), the approximate solution 
(\ref{uapso}) can be rewritten as
\begin{equation}
u\approx\frac{r_s^2}{2L^2}\,[1+\delta\cos(\vp-\sigma\vp)]\,,
\label{eq:shif3}
\end{equation}
where $\sigma~(\ll 1)$ is defined by 
\begin{equation}
\sigma \equiv \frac32\,\frac{1}{(1+\delta)\rho_0}
-\frac12\,\ve\,c_1\,p\,(p+1)\,(1+\delta)\,\rho_0^{p+1}\,.
\label{eq:shif4}
\end{equation}

{}From Eq.~(\ref{eq:shif3}) we find that in one orbit the angle between two perihelia 
is larger than $2\pi$ approximately by $2\pi\sigma$, or
\begin{equation}
\Delta\vp-2\pi=
\frac{3\pi}{(1+\delta)\rho_0}-\pi\,\ve\,c_1\,p\,(p+1)\,(1+\delta)\,\rho_0^{p+1}
\approx\frac{3\pi}{(1+\delta)\rho_0}\left[1-\frac13\,c_1\,\ve\,p(p+1)
(1+2\delta)\rho_0^{p+2}\right]\,.
\label{eq:shif5}
\end{equation}
The experimental bound on the shift $\Delta \vp-2\pi$ for Earth, 
based on several thousands of optical observations 
of planetary positions, is \cite{Mercury} 
\begin{equation}
\Delta\varphi-2\pi=5\pm1~~\textrm{arcsec/century}.
\label{delvpcon}
\end{equation}
For Mercury the bound is $43.1\pm0.1$ arcsec/century.
Since the GR contribution is given by
\begin{equation}
\label{eq:PERGR}
\Delta\vp_{\rm GR}=\frac{3\pi}{(1+\delta)\rho_0}\, ,
\end{equation}
one sets the modifications of gravity to contribute less than 
the experimental relative error, that is
\begin{equation}
\left\vert\frac{\Delta\vp_\ve}{\Delta\vp_{\rm GR}}\right\vert
=\frac13\,|c_1|\,\ve\,p(p+1)
(1+2\delta)\rho_0^{p+2}<\frac{1}{5}\,.
\label{delvpco2}
\end{equation}

The correction term remains as such if
\begin{equation}
\label{eq:shifCR}
\ve \rho_0^{p+2} \ll 1\,.
\end{equation}
This condition is not necessarily satisfied in the whole domain 
of convergence $D$. 
Therefore, together with the bound (\ref{delvpco2}), 
this can be used to constrain modified gravity models.

\subsection{Retardation of light}

Let us study the gravitational time delay effect in light signals. 
For a light signal propagating from $\rho_0$ to $\rho$, 
the integration of Eq.~(\ref{eq:luce1}) with respect to $\rho$ gives
\begin{equation}
t=r_s\int_{\rho_0}^{\rho}\de \bar{\rho} \left[AB\left(1
-\frac{A}{A_0}\frac{\rho_0^2}{\bar{\rho}^2}\right)\right]^{-1/2}\,.
\label{eq:cass2}
\end{equation}
Expanding the integrand in $\ve$ and assuming that both $\rho_0$ and $\rho$ are 
much greater than unity, the integral (\ref{eq:cass2}) is approximately given by
\begin{equation}
t(\rho_0,\rho)\approx
\frac{r_s\sqrt{\rho_0-1}}{\sqrt{\rho_0}}
\int_{\rho_0}^\rho\de \bar{\rho}\,{\frac {{\bar{\rho}}^{2}}
{\left( \bar{\rho}-1\right) \sqrt {{\bar{\rho}}^{2}
-{\rho_0}^{2}}}}
-\frac{\ve r_s}2\int_{\rho_0}^\rho \de \bar{\rho}\,
\left\{\rho\,\frac{\bar{\rho}^p \left[ \bar{\rho}^2(c_1+c_2)
-\rho_0^2 (2c_1+c_2)\right]+c_1\rho_0^{p+2}}
 {\left(\bar{\rho}^2-\rho_0^2\right)^{3/2}}
 +{\cal O}(\bar{\rho}^{p-1})\right\}\,,
\label{tcas}
\end{equation}
where $\rho$ represents the position of the satellite. 
See Ref.~\cite{Asada} for the similar calculation about the 
gravitational time-delay effect.

The second term in Eq.~(\ref{tcas}) corresponds to the $\ve$ contribution, $t_{\ve}$.
Under the condition $\rho\gg\rho_0$, the $\ve$ contribution can be 
evaluated as
\begin{eqnarray}
& & p=1,\qquad r_s^{-1}t_{\ve} (\rho_0,\rho) \approx-\frac14\,(c_1+c_2)\ve\rho^2
+\frac18\,\ve\,\rho_0^2\,[(c_2-3c_1)+2(c_1-c_2)\ln(2\rho/\rho_0)]\,,\\
& & p=2,\qquad r_s^{-1}t_{\ve} (\rho_0,\rho)\approx-\frac16\,(c_1+c_2)\ve\rho^3
+\frac14\,(c_1-c_2)\,\ve\,\rho_0^2\rho+\frac14\,\bar c\,\ve\,\rho_0^2\,,\\
& & p=3,\qquad r_s^{-1}t_{\ve} (\rho_0,\rho)\approx-\frac18\,(c_1+c_2)\ve\rho^4
+\frac18\,(c_1-c_2)\,\ve\,\rho_0^2\rho^2-\frac{1}{32}\,\bar c\,\ve\,\rho_0^4\,
[13-12\ln(2\rho/\rho_0)]\,,\\
& & p \geq 4,\qquad r_s^{-1}t_{\ve} (\rho_0,\rho) \approx-\frac12\,
\frac{c_1+c_2}{p+1}\,\ve\rho^{p+1}+\frac14\,\frac{c_1-c_2}{p-1}\,\ve\,
\rho_0^2\rho^{p-1}
-\frac{1}{4(p-3)}\,\bar c\,\ve\,\rho_0^3\,\rho^{p-3}\,,
\end{eqnarray}
where we have introduced the constant $\bar c$ 
defined as $c_1$ when $c_1=c_2$.

If $p \geq 4$ the time difference between two points $\rho_1$ and $\rho_2$ 
coming from the $\ve$ contribution is  
\begin{equation}
\Delta t_{\ve} \approx -\frac{r_s}2\,\frac{c_1+c_2}{p+1}\,
\ve\,(\rho_1^{p+1}+\rho_2^{p+1})\,.
\label{eq:retL4}
\end{equation}
The two contributions add because in the first integral one has
propagation from the satellite to Sun ($\de t/\de \rho<0$), and
in the other one from Sun to Earth ($\de t/\de \rho>0$).
Since the standard GR contribution to Eq.~(\ref{tcas}) is $t_{\rm GR} \approx
r_s\ln\!\left[ 2\rho/\rho_0 \right]$, the time difference 
between two points $\rho_1$ and $\rho_2$ can be estimated as
\begin{equation}
\Delta t_{\rm GR} \approx
r_s\ln\!\left[ \frac{4\rho_1\rho_2}{\rho_0^2} \right]\,.
\label{eq:retL1}
\end{equation}
The ratio among the two contributions is then given by 
\begin{equation}
\label{eq:ROL1}
\frac{\Delta t_\ve}{\Delta t_{\rm GR}}=-\frac12\,\ve\,
\frac{c_1+c_2}{p+1}\,\frac{\rho_1^{p+1}
+\rho_2^{p+1}}{\ln(4\rho_1\rho_2/\rho_0^2)}\, .
\end{equation}
The bound regarding the ratio between the measured delay and the one 
predicted by GR comes from the Viking mission on Mars, which gives the result 
$\Delta t_{\rm exp}/\Delta t_{\rm GR}=1.000\pm0.001$ \cite{Viking}.
Hence this gives the bound
\begin{equation}
\left| \frac{\Delta t_\ve}{\Delta t_{\rm GR}} \right|<10^{-3}\,.
\label{retarcon}
\end{equation}
Setting $\rho_1 \approx \rho_2 \approx \rho$, this condition translates into 
\begin{equation}
\frac{\ve |c_1+c_2|}{2(p+1)} \frac{\rho^{p+1}}{\ln (2\rho/\rho_0)}<10^{-3}\,.
\end{equation}
This is generally satisfied in the domain of convergence, 
as the logarithmic term grows slowly with $\rho$.

\subsection{Gravitational Redshift and Equivalence Principle}

Let us finally consider the gravitational redshift. 
In this case, for a light signal propagating at different heights $r$ and $r_1$, 
the ratio of the frequencies $\nu$ and $\nu_1$ for corresponding heights
is given by 
\begin{equation}
\label{eq:gravred}
\frac{\nu}{\nu_1}=\sqrt{\frac{A(r)}{A(r_1)}}\approx1+\frac12\,(\rho_1^{-1}-\rho^{-1})
+\frac12\,c_1\,\ve\,(\rho^p-\rho_1^p)\,.
\end{equation}
The $\ve$-dependent term is much smaller than the standard GR one in the $D$ domain.
Defining $\Delta\nu=\nu-\nu_1$, it then follows that
\begin{equation}
\frac{\Delta\nu_\ve/\nu}{\Delta\nu_{\rm GR}/\nu}=
\frac{c_1\ve\,\rho\,\rho_1(\rho^p-\rho_1^p)}{\rho-\rho_1}\,.
\label{eq:grd1}
\end{equation}
The bound on $(\Delta\nu_\ve/\nu)/(\Delta\nu_{\rm GR}/\nu)$ comes from 
the experiment of an hydrogen-maser clock on a rocket launched to 
an altitude of about $10^{7}$ m \cite{Vessot}, which corresponds to 
$\Delta\nu_{\rm exp}/\Delta\nu_{\rm GR}=1\pm0.0002$.
This leads to the following bound
\begin{equation}
\frac{\Delta\nu_\ve/\nu}{\Delta\nu_{\rm GR}/\nu}<2 \times 10^{-4}\,.
\label{gracon}
\end{equation}

In the non-relativistic limit, the gravitational potential $V$ can be identified as $g_{00}=1+2V$, that is
\begin{equation}
V=-\frac1{2\rho}+\frac12\,c_1\ve\,\rho^p=-\frac{G_NM_{\odot}}{r}
+\frac12\,c_1\,\ve\left(\frac r{2G_NM_{\odot}}\right)^p\,.
\label{eq:EQPR}
\end{equation}
This only depends on the mass $M_{\odot}$ of Sun, not on the mass/properties
of the test particle.  Hence all test particles with same distance
from the center will feel the same acceleration and the equivalence
principle will not be violated.

\section{Power-law $f(\G)$ model}

The approach we have used so far works only if the iterative parameter $\ve$
is much smaller than 1, and the method works better for smaller $\ve$.  
As a result this method cannot be evidently applied to
all forms of $R+f(\G)$. 
For example, let us consider the simple power-law case \cite{Davis}
\begin{equation}
\label{eq:polinom}
f(\G)=\lambda\,\sqrt{\G_*} \left(\frac{\G^2}
{\G_*^2}\right)^k\,,
\end{equation}
such that $f(\G)$ is defined for all real values of $k$ and $\G$.
In fact, for the spherically symmetric spacetime, 
this Lagrangian will give rise to terms typically of order
\begin{align}
\G\,f_{,\G}-f&=\lambda\,(2k-1)\,\bar\ve\sqrt{\G_s}\,
\left[\left(\frac{\G}{\G_s}\right)^{\!2}\right]^k \,,\\
\G_s^2\,f_{,\G\G}&=2\,\lambda\,k\,(2k-1)\,\bar\ve\sqrt{\G_s}\,
\left[\left(\frac{\G}{\G_s}\right)^{\!2}\right]^{k-1} \,,
\label{eq:pospow}
\end{align}
where
\begin{equation}
\label{eq:pow1}
\bar\ve\equiv\ve^{1-4k}=\left[ \sqrt{\G_*/\G_s} \right]^{1-4k}\,.
\end{equation}

The fact that $\bar\ve$ can be larger than one, implies that the
corrections to the \swa\ metric may become large unless $1-4k>0$, that
is $k<1/4$. The GR case corresponds to $k=1/2$, 
in which the contribution of the GB term vanishes.
If $k\neq1/2$ but close to it, then we still require that $\lambda |2k-1|/\ve\ll1$ 
in order to regard these terms as corrections to the equations of motion.
In general, if $k\neq1/2$ and $k>1/4$, the iterative method cannot 
be used to find approximate solutions to the equations of motion. 
In such cases the solutions need to be obtained by numerical integrations.

If $k<1/4$ then $f_{,\G\G}$ is proportional to $[(\G_s/\G)^2]^{1-k}$,
so that this term blows up as $\G \to 0$. 
In cosmological backgrounds the GB term $\G=24H^2(H^2+\dot{H})$
changes sign from negative to positive during the transition from
the matter era to the accelerated epoch \cite{CFG}.  This leads to the
divergence of $f_{,\G \G}$ at $\G=0$, which means that the pure
power-law $f(\G)$ model is not cosmologically viable.  We also note
that the GB term inside and outside a spherically symmetric body (mass
$M_{\odot}$ and radius $r_{\odot}$) with homogeneous density are given
by $\G=-48(G_N M_{\odot})^2/r_{\odot}^6$ and $\G=48(G_N
M_{\odot})^2/r^6$, respectively.
As we move from the interior to the exterior of the star the
GB term also crosses 0 from negative to positive. Although it is
possible to derive iterative spherically symmetric solutions for
$k<1/4$ by using the expansion in terms of $\bar{\epsilon}$, the
power-law $f(\G)$ model is out of our interest because of the problems
mentioned above.

\section{Model A}
\label{modelasec}

In this section we study the model (A) given in Eq.~(\ref{modela}), i.e.
\begin{equation}
f(\G)= \ve \left[ -\alpha \lambda \sqrt{\G_s} + \lambda 
\sqrt{\G_s} g (x) \right]\,,
\label{modelA}
\end{equation}
where
\begin{equation}
g(x) =x \arctan x - \frac{1}{2} \ln (1 + x^2)
\qquad{\rm and}\qquad x = \frac{\G}{\G_s \varepsilon^2}\,. 
\end{equation}
The Lagrangian is function of $\G/\G_*$, but we choose to write it in this form
so that the dependence on $\ve$ becomes explicit. 
Since $f_{,\G \G}>0$ for positive $\lambda$ and $\G_*$, 
there is no singularity of this quantity unlike the power-law
$f(\G)$ model.

Let us discuss then the different contributions to  the equations of motion (\ref{eq:VC2}).
For the model (\ref{modelA}) we have
\begin{align}
\G f_{,\G} - f &=  \G \frac{\lambda}{\sqrt{\G_s} \varepsilon} \arctan x + \alpha
\lambda \varepsilon \sqrt{\G_s} - \lambda \varepsilon \sqrt{\G_s} g (x) \nonumber \\
&= \ve \left[ \alpha \lambda \sqrt{\G_s}
+\frac12 \lambda \sqrt{\G_s}
\ln\! \left( 1 + \frac{\G^2}{\G_s^2 \varepsilon^4} \right) \right]\,,
\end{align}
which is of the order of $\ve \alpha \lambda \sqrt{\G_s}$ plus a logarithmic correction.
The other terms that appear in the equations of motion, i.e.\ $f_{,\G\G}$ and $f_{,\G\G\G}$, 
can be written as follows
\begin{align}
  f_{,\G\G} & =  \frac{\lambda \sqrt{\G_s}}{\G_s^2 \varepsilon^3} \frac{\de^2
  g}{\de x^2} = \frac{\lambda \sqrt{\G_s} \varepsilon}{\G_s^2 \varepsilon^4
  + \G^2},\\
  f_{,\G\G\G} & = \frac{\lambda \sqrt{\G_s}}{\G_s^3 \varepsilon^5} \frac{\de^3
  g}{\de x^3} = - \frac{2 \lambda \sqrt{\G_s} \G \varepsilon}{(\G_s^2
  \varepsilon^4 + \G^2)^2}\,,
\end{align}
which are both of the order of ${\cal O} (\varepsilon)$ 
(as typically $|\G| \gg \G_s$ in the solar system). 

We write the metric in terms of the expansion parameter $\ve$:
\begin{equation}
\de s^2 = - \left[ 1 - \frac{r_s}{r} + \varepsilon\phi_1(r)
 + \varepsilon^2\phi_2(r) \right] \de t^2+ \left[ 1 - \frac{r_s}{r} + \varepsilon\psi_1 (r)
+ \varepsilon^2\psi_2 (r) \right]^{- 1} \de r^2 
+ r^2 (\de \theta^2 + \sin^2
\theta\, \de \varphi^2)\,.
\label{permetric}
\end{equation}
Although the second-order contribution is not used in order to
obtain experimental bounds on the model, we will evaluate it 
to check whether the terms in the series become smaller
for higher orders of $\ve$ and to verify that $D_2\subseteq D$. 
Because of the Bianchi identities we can use only two equations, 
i.e.\ the 0-0 and the 1-1 equations, 
as the others are automatically satisfied. 
The second-order equations follow after solving the 
first-order equations.

\subsection{Spherically symmetric solutions and the domain of convergence}

Linearizing the 0-0 component of the modified Einstein equations at first-order in $\ve$, 
we obtain the differential equation for $\psi_1$ in terms of $\rho=r / r_s$:
\begin{equation}
\rho \frac{\de\psi_1}{\de\rho} + \psi_1 =32\sqrt{3}\lambda \rho^3
+12\sqrt{3}\lambda \rho^2\ln(\rho)
+ (4\ln\ve-2\alpha-28)\sqrt{3}\lambda \rho^2\,, 
\end{equation}
whose particular solution is
\begin{equation}
\psi_1 = 8\sqrt{3}\lambda\rho^3
+4\sqrt{3}\lambda\rho^2\ln\rho
+\frac23\sqrt{3}\left( 2\ln\ve-\alpha-16 \right )
\lambda\rho^2\,.
\label{psi1}
\end{equation}
Here we have neglected the contribution coming from the homogeneous solution, 
as this would correspond to an order $\ve$ renormalization contribution to the mass 
of the system. Although $\ve\ll1$ the term in $\ln\ve$ only contributes by a factor 
of order $10^2$. Hence the first term on the r.h.s.\ of Eq.~(\ref{psi1}) 
dominates over the last term.

The 1-1 component of the Einstein equations gives the equation 
used to determine $\phi$ as follows
\begin{equation}
(\rho-\rho^2)\frac{\de\phi_1}{\de\rho}+\phi_1
=8\sqrt{3}\lambda\rho^4
-2\sqrt{3} (10+6\ln\rho+2\ln\ve-\alpha) \lambda\rho^3
-2\sqrt{3}( \alpha-6\ln\rho-2\ln\ve-6)\lambda\rho^2+\rho\psi_1\,.
\label{phi1}
\end{equation}
Substituting the solution (\ref{psi1}) into Eq.~(\ref{phi1}), 
we get the following solution for $\phi_1$:
\begin{equation} 
\phi_1=-\frac{16}3\sqrt{3}\lambda\rho^3
+\frac23\sqrt{3} \left( 4-\alpha+6\ln\rho+2\ln \ve 
\right)\lambda\rho^2\,.
\end{equation}
Since $\rho\gg1$, the largest contributions to $\psi_1$ and
$\phi_1$ correspond to the ones proportional to $\rho^3$, 
which are different from the \swa-de Sitter contribution (which grows as $\rho^2$). 
Hence the model (\ref{modela}) gives rise to the corrections
larger than that in the cosmological constant case by a factor $\rho$. 

In the Appendix we present the equations for the second-order quantities 
$\psi_2$ and $\phi_2$. Discarding the homogeneous part of 
solutions of Eqs.~(\ref{appen1}) and (\ref{appen2}), we obtain
\begin{eqnarray}
\hspace*{-3.3em}& & \psi_2=-512\,{\lambda}^{2}{\rho}^{7}-\frac {4}{21} \lambda^{2} 
\left( -3972+168\,\ln\ve +504\,\ln \rho 
-84\,\alpha \right) {\rho}^{6}-\frac {4}{21}\,{\lambda}^{2} \left( -
322\,\ln \ve +161\,\alpha-966\,\ln\rho+1547 \right) {\rho}^{5},\label{psi2} 
\label{psi2eq}\\
\hspace*{-3.3em}& & \phi_2={\frac {1216}{7}}\,{\lambda}^{2}{\rho}^{7}
-\frac{4}{21}\lambda^2 \left( -14\,\alpha+28\,\ln\ve+1014
+84\,\ln\rho \right) {\rho}^{6}-{\frac {4}{21}}\,{\lambda}^{2
} \left( 294\,\ln \rho -399-49\,\alpha+98\,\ln\ve\right) {\rho}^{5},
\label{phi2eq} 
\end{eqnarray}
where the dominant terms are the first terms in Eqs.~(\ref{psi2eq}) and (\ref{phi2eq}).

So far we have found the solutions up to the order of $\ve^2$. 
This expansion is meaningful if each term of the expansion is smaller than the previous one. 
For the solar system experiments, we will consider the case $\rho\gg1$. 
This automatically imposes that the \swa\ contribution $r_s/r$ is smaller than the Minkowski
value 1, which corresponds to the weak field approximation. 
The first-order correction in $\ve$ is smaller than 
the \swa\ contribution if
\begin{equation}
\ve|\psi_1| \ll 1/\rho,\qquad {\rm and} \qquad
\ve|\phi_1| \ll 1/\rho\,.
\label{psiphicon}
\end{equation}
Since the dominant terms in $\psi_1$ and $\phi_1$ have the dependence 
$\lambda \rho^3$, the conditions (\ref{psiphicon}) translate into
\begin{equation}
\rho\ll|\lambda\ve|^{-1/4} .
\label{rhovecon}
\end{equation}
Therefore, at first order, the domain of convergence is
\begin{equation}
1\ll\rho\ll 10^{11}\,|\lambda|^{-1/4}\,.
\end{equation}
The consistency of this inequality requires that $|\lambda| \ll 10^{44}$. 
The second-order terms in $\ve$ can be neglected 
if the following conditions are satisfied in the same region 
\begin{equation}
\ve^2\psi_2\ll\ve\psi_1\qquad{\rm and}\qquad
\ve^2\phi_2\ll\ve\phi_1\,.
\label{psi2con}
\end{equation}
Since the dominant contributions in $\psi_2$ and $\phi_2$
have the dependence $\lambda^2 \rho^7$, the conditions
(\ref{psi2con}) are equivalent to the requirement (\ref{rhovecon}).
Hence our solutions using the expansion in $\ve$ can be justified
in the domain of convergence $D$.

\subsection{Solar system constraints}

We have seen that in the domain $D$ of convergence the second-order term 
proportional to $\ve^2$ can be neglected. 
Moreover, in this domain, we can approximate the
solutions further, keeping only the highest power in $\rho=r/r_s$ 
at the order $\ve$. Hence we have, for local gravity considerations, 
the metric (\ref{eq:genmet}) with 
\begin{eqnarray}
& &A=1-\frac{1}{\rho}-\frac{16}{3}\sqrt{3}\lambda
\ve\rho^3\,[1+{\cal O}(\ve,\rho^{-1})]\, ,\\
& &B=1-\frac{1}{\rho}+8\sqrt{3}\lambda\ve
\rho^3\,[1+{\cal O}(\ve,\rho^{-1})]\,.
\end{eqnarray}
Therefore, in the domain of convergence, we will always regard the order 
$\ve$ quantity as a correction to the \swa\ contribution, 
and for the order $\ve$ we will always keep only the highest power in $\rho$.
Since the solution given above is an approximate one
valid in the domain of convergence, it does not make sense to see
whether or not this metric is asymptotically flat: this would
correspond to the solution in a region outside the domain of convergence. 
However, even having the real solution at hand (which is
not the case), one should not trust it far away from Sun, 
as other forces would provide large contributions.

In the following let us place constraints on the model parameter
$\lambda$ by using a number of experimental bounds 
discussed in Sec.~\ref{solarconst}.
\begin{itemize}
\item (A) Deflection of light

{}The constraint (\ref{eq:Lct}) coming from the deflection of light gives
\begin{equation}
\frac{|c_1-c_2|}{2(p-1)} \ve \rho_0^2 \rho_{\rm Earth}^{p-1}<10^{-4}\,.
\end{equation}
For this model we have that
\begin{equation}
\label{eq:c1c2mod1}
c_1=-\frac{16}{3}\sqrt{3}\lambda\,,\quad
c_2=8\sqrt{3}\lambda\,,\quad p=3\,.
\end{equation}
These numbers define the model and are the same for all the remaining constraints.
The radius of Sun in units of the \swa\ radius is $\rho_0=2.35\times10^5$. 
The distance of Earth from Sun in units of the \swa\ radius is 
$\rho_{\rm Earth}=5.08\times10^7$.
This translates into the following bound
\begin{equation}
\lambda < 1 \times 10^{15}\,,
\label{cona1}
\end{equation}
where we have used the value (\ref{epvalue}) for $\ve$.

\item (B) Cassini experiment

{}As we have already seen, the Cassini experiment places the bound $y_\ve<10^{-14}$, i.e.
\begin{equation} 
v_{\rm Earth} \frac{|c_1-c_2|}{p-1}\,\ve\rho_0(\rho_{\rm Cass}^{p-1}
+\rho_{\rm Earth}^{p-1})<10^{-14}\,.
\end{equation}
The speed of Earth in units of the speed of light is $v_{\rm Earth}=9.93\times10^{-5}$,
whereas the distance of Saturn in units of the \swa\ radius
 is $\rho_{\rm Cass}=4.85\times10^8$.
This gives 
\begin{equation} 
\lambda < 2 \times 10^{12}\,.
\label{cona2}
\end{equation}
\item (C) Perihelion shift

The bound (\ref{delvpco2}) coming from the shift of the perihelion 
of Earth leads to 
\begin{equation}
\frac13\,|c_1|\,\ve\,p(p+1)
(1+2\delta_{\rm Earth})\rho_{\rm Earth}^{p+2}
<\frac{1}{5}\,.
\label{peribound}
\end{equation}
For the eccentricity $\delta_{\rm Earth}=0.02$ and the perihelion of Earth
in units of the \swa\ radius $\rho_{\rm Earth}=4.98\times10^7$, we obtain 
the constraint 
\begin{equation} 
\lambda< 2 \times10^{5}\,.
\label{cona3}
\end{equation}
For Mercury, $\delta_{\rm Hg}=0.2$ and  $\rho_{\rm Hg}=1.56\times10^7$, the bound is slightly weaker, $\lambda<5\times10^5$.

\item (D) Retardation of light

The bound (\ref{retarcon}) coming from the retardation of light,
together with Eq.~(\ref{eq:retL1}), gives the following constraint
\begin{equation}
\frac12\,\ve\,
\frac{|c_1+c_2|}{p+1}\,\frac{\rho_{\rm Mars}^{p+1}+\rho_{\rm Earth}^{p+1}}
{\ln(4\rho_{\rm Mars}\rho_{\rm Earth}/\rho_0^2)} <10^{-3}\,,
\label{eq:MA1}
\end{equation}
which gives, with $\rho_{\rm Mars}=7.71\times10^7$ in units of the \swa\ radius,
\begin{equation} 
\lambda<5 \times 10^{12}\,.
\label{cona4}
\end{equation}

\item (E) Gravitational redshift

{}From the constraint (\ref{gracon}) coming from the gravitational
redshift, together with Eq.~(\ref{eq:grd1}), it follows that 
\begin{equation} 
\frac{|c_1|\ve_{\rm Earth}\,\rho_2\,\rho_1(\rho_2^p-\rho_1^p)}
{\rho_2-\rho_1}<2 \times 10^{-4}\,.
\label{graredcon}
\end{equation}
Here $\ve_{\rm Earth} \approx 10^{-57}$, $\rho_1$ is the radius of
Earth in units of its \swa\ radius, i.e.\ $\rho_1=7.18\times10^8$, and
$\rho_2=1.84\times10^9$ is the distance of the experimental 
apparatus (for a height of 10$^4$ km).
We then obtain the following bound
\begin{equation} 
\lambda<3 \times 10^{15}\,.
\label{cona5}
\end{equation}
\end{itemize}

The tightest constraint on $\lambda$ comes from the perihelion shift 
experiment. This bound is weak so that the $f(\G)$ model (\ref{modela})
can be consistent with solar system constraints for a wide range of 
the model parameter.

\section{Model B}
\label{modelbsec}

Let us next proceed to the constraints on the model (\ref{modelb}).
This model can be written as
\begin{equation} 
f (\G) = \ve \left[ -\alpha \lambda \sqrt{\G_s} + 
\lambda \sqrt{\G_s} g (x) \right]\,,
\end{equation} 
where
\begin{equation} 
g (x) = x \arctan x, \qquad\mathrm{and}\qquad x 
= \frac{\G}{\G_s \ve^2}\,.
\end{equation} 
We shall derive vacuum solutions for the spherically symmetric 
metric (\ref{eq:genmet}) by using the same expansion parameter
$\ve$ defined in (\ref{vedef}).
The term $\G f_{,\G} - f$ in the equations of motion 
can be estimated as
\begin{equation} 
\G f_{,\G} - f = \ve \lambda \sqrt{\G_s} 
\left[ \alpha + \frac{\G^2}{\G_s^2 \ve^4 + \G^2} \right]\,. 
\end{equation} 
Since $\G^2 \gg \G_s^2 \ve^4$, it follows that
\begin{equation} 
\G f_{,\G} - f \approx \ve  \lambda \sqrt{\G_s} (\alpha+1)
+{\cal O} (\ve^5)\,,
\end{equation} 
which works as a cosmological constant at lowest order.
Note that we have $f_{,\G\G} \sim \varepsilon^5+{\cal O}(\ve^9)$ and
$f_{,\G\G\G} \sim \varepsilon^{5}+{\cal O}(\ve^9)$, 
so that these terms are higher than the linear order in $\ve$.
These properties are different from those in the model (\ref{modela}).

For the model (\ref{modelb}) the dominant contribution to the Schwarzschild
metric comes from the linear term in $\ve$ and the next order corrections 
correspond to terms in $\ve^5$. Hence we look for a metric of the form 
\begin{equation} 
\de s^2= -\left[ 1 - \frac{r_s}{r}+\ve \phi_1(r)+\ve^5 \phi_2(r) 
\right] \de t^2 +\left[1-\frac{r_s}{r}+\ve \psi_1(r) 
+\ve^5 \psi_2 (r) \right]^{-1} \de r^2+ 
 r^2 (\de \theta^2 + \sin^2 \theta\, \de \varphi^2)\,.
 \label{ep5so}
\end{equation} 
Linearizing the 00 component of the Einstein equation (\ref{geeq})
at first order in $\ve$, we obtain the following differential equation 
\begin{equation} 
\rho \frac{\de\psi_1}{\de\rho} - \psi_1+
2 \sqrt{3} \lambda (1 + \alpha) \rho^2 = 0\,,
\end{equation} 
which has the particular solution
\begin{equation} 
\psi_1 = - \frac{2 \sqrt{3}}{3} \lambda (1 + \alpha) \rho^2\,.
\end{equation} 
The 11 component of the linearized Einstein equation gives 
\begin{equation} 
(\rho^2 - \rho) \frac{\de\phi_1}{\de\rho} - \phi_1 + 
\frac{4 \sqrt{3}}{3} \lambda (\alpha +1) \rho^3 
-2 \sqrt{3} \lambda (\alpha + 1) \rho^2 = 0\,,
\end{equation} 
whose particular solution is 
\begin{equation} 
\phi_1 = - \frac{2 \sqrt{3}}{3} \lambda (1 + \alpha) \rho^2\,.
\end{equation} 
These expressions for the metric corrections represent the contribution of an
effective cosmological constant (as in the case of Schwarzschild de Sitter metric). 
This solves the equations of motion up to the order $\varepsilon$.

Let us derive next-order solutions $\psi_2(r)$ and $\phi_2(r)$.
The differential equations for $\psi_2$ and $\phi_2$ are
\begin{eqnarray}
& &\rho \frac{\de\psi_2}{\de\rho} + \psi_2 - 2 \sqrt{3} \lambda \rho^{14} (128 \rho - 123) =0\,,\\
& &(\rho^2 - \rho) \frac{\de\phi_2}{\de\rho} - \phi_2 + \frac{2 \sqrt{3}}{5} \lambda \rho^{14}
(80 \rho^2 - 146 \rho + 65) = 0\,,
\end{eqnarray}
which have the following particular solutions
\begin{eqnarray}
& & \psi_2=\frac{2 \sqrt{3}}{5} \lambda \rho^{14} (40 \rho - 41)\,,\\
& & \phi_2 = - \frac{2 \sqrt{3}}{15} \lambda \rho^{14} (16 \rho-13)\,.
\end{eqnarray}
In principle one can repeat this method order by order in
$\ve$. It is then clear that the vacuum solution will not be
strongly constrained by solar system experiments, 
as the first non-zero contribution will be that of a cosmological constant. 
To be more precise, the domain of convergence $D$ is defined as
\begin{equation}
\label{eq:mb1}
\ve\lambda(1+\alpha)\rho^2\ll 1/\rho \ll 1\,,
\end{equation}
which implies that
\begin{equation}
\label{eq:mb2}
1\ll\rho\ll \frac{\ve^{-1/3}}{[\lambda(1+\alpha)]^{1/3}}
 \approx \frac{10^{15}}{[\lambda(1+\alpha)]^{1/3}}\,.
\end{equation}
For the consistency of this inequality we require that 
$\lambda (1+\alpha) \ll 10^{45}$.
If $\lambda$ and $\alpha$ are of the order of unity, the domain $D$
corresponds to the distance $10^5$\,cm$\,\lesssim r \lesssim 10^{20}$\,cm.

The contributions $\ve^5 \psi_2$ and $\ve^5 \phi_2$ can be negligible 
relative to the first-order contributions provided that
$\lambda\ve^5\rho^{15}\ll\lambda(1+\alpha)\ve\rho^2$, i.e.
\begin{equation}
\rho \ll (1+\alpha)^{1/13}\ve^{-4/13}\,.
\label{eq:mbe2b}
\end{equation}
If $\alpha={\cal O}(1)$ the domain $D_2$ of convergence corresponds to
$r \lesssim 10^{19}$\,cm.  One may wonder if this trend continues at
next order, that is, $\ve^6$. When the previous two contributions are
coupled, this gives rise to terms of the order $\lambda^2(1+\alpha)\ve^6$. 
Introducing the corrections $\ve^6\,\psi_3$ and $\ve^6\,\phi_3$ 
to the metric and expanding the equations of motion at order $\ve^6$, 
we obtain the following differential equations
\begin{eqnarray}
& & \rho\frac{\de\psi_3}{\de\rho}+\psi_3-\lambda^2\,(\alpha+1)\,\rho^{17}\,(608\rho-1136)=0\,,\\
& & (\rho^2-\rho)\frac{\de\phi_3}{\de\rho}-\phi_3-\lambda^2(\alpha+1)\,\rho^{17}
\left( 32\rho^2+\frac{1064}{45}\rho-\frac{304}5 \right)=0\,,
\end{eqnarray}
whose solutions are given by 
\begin{eqnarray}
& &\psi_3=\lambda^2(1+\alpha)\,\rho^{17}\,\left( 32\rho-\frac{568}{9} \right)\,,\\
& &\phi_3=\lambda^2(1+\alpha)\,\rho^{17}\,\left( \frac{16}{9}\rho
+\frac{152}{45} \right)\,.
\end{eqnarray}
Therefore the expansion is meaningful for
$\lambda(1+\alpha)\ve^6\rho^{18}\ll\ve^5\rho^{15}$, that is,
$D_3=D\subseteq D_2$. 
This shows that the domain of convergence at the order $\ve^6$
coincides with $D_2$.

Let us discuss solar system constraints on the model (\ref{modelb}) 
by using the experimental bounds 
discussed in Sec.~\ref{solarconst}.
We will show that the strongest
bound comes from the shift of the perihelion of Earth.
\begin{itemize}
\item (A) Deflection of light

For the model (\ref{modelb}) we have that 
\begin{equation}
c_1=c_2=- \frac{2 \sqrt{3}}{3} \lambda (1 + \alpha)\,,\quad p=2\,.
 \label{eq:c1c2mod1d}
\end{equation}
Since $c_1=c_2$, the constraint (\ref{eq:Lct}) coming from
the deflection of light is trivially satisfied.

\item (B) Cassini experiment

Since $c_1=c_2$ at order $\ve$, the bound (\ref{cassibo}) 
of the Cassini experiment is fulfilled.

\item (C) Perihelion shift

The bound coming from the shift of the perihelion of Earth
corresponds to (\ref{peribound}) with 
$\delta_{\rm Earth}=0.02$ and $\rho_{\rm Earth}=4.98\times10^7$.
This leads to
\begin{equation} 
\lambda(1+\alpha)<1 \times10^{14}\,.
\label{cona3d}
\end{equation}
From the bound of Mercury we obtain a similar constraint.

\item (D) Retardation of light

Using the bound (\ref{eq:MA1}) coming from the retardation of light
with $\rho_{\rm Mars}=7.71\times10^7$, 
we obtain the following constraint
\begin{equation} 
\lambda(1+\alpha)<6 \times10^{20}\,.
\label{cona4d}
\end{equation}

\item (E) Gravitational redshift

Using the constraint (\ref{graredcon}) of the gravitational redshift 
with $\ve_{\rm Earth}=10^{-57}$, $\rho_1=7.18\times10^8$, 
and $\rho_2=1.84\times10^9$, it follows that
\begin{equation} 
\lambda(1+\alpha)<3 \times 10^{25}\,.
\label{cona5d}
\end{equation}
\end{itemize}

The bottom line is that the model B, having $p=2$, is less constrained than the model A.

\section{Conclusions}

In this paper we have discussed solar system constraints 
on $f(\G)$ gravity models that are cosmologically viable.
These models give rise to power-law corrections of the form 
$(r/r_s)^p$ to the \swa\ metric, which are characterized by an 
expansion parameter $\ve=\sqrt{\G_*/\G_s} \approx 10^{-46}$ for Sun.
The smallness of this parameter allows us to find approximate 
vacuum solutions in a spherically symmetric spacetime. 

In order to confront $f(\G)$ models with a number of solar-system experiments,
we have carried out general analysis for estimating their deviation from 
General Relativity. These include the experiments such as deflection of light, 
Cassini tracking, perihelion shift of Earth, retardation of light, and gravitational redshift.
The results we have derived can be generally applied to any 
modified gravity models which have power-law corrections 
to the \swa\ metric.

The $f(\G)$ models given in Eqs.~(\ref{modela}) and (\ref{modelb}) are
designed to give rise to a late-time cosmic acceleration preceded by a
matter-dominated epoch.  We find that these models can satisfy all of
solar system constraints discussed in literature for a wide range of
model parameters.  For the model (\ref{modela}) there exists a
logarithmic correction $(\lambda/2)\sqrt{\G_*} \ln (\G^2/\G_*^2)$ to
the Lagrangian density in the region of high density ($\G^2 \gg
\G_*^2$).  The tightest bound comes from the shift of perihelion of
Earth, but the constraint on the parameter $\lambda$ is weak:
$\lambda < 2 \times 10^5$. 
In order to set stronger bounds on
these theories, it is then necessary to have better measurement of
the quadrupole moment of Sun, as it affects the perihelion
shift.  For the model (\ref{modelb}) the leading correction term to
the Lagrangian density corresponds to
$(\lambda/3)\sqrt{\G_*}\,\G_*^2/\G^2$, whose effect is very small even
compared to the model (\ref{modela}).  Hence the model parameter for
the model (\ref{modelb}) is very weakly constrained: $\lambda
(1+\alpha) < 10^{14}$.

The main reason why the $f(\G)$ models can satisfy solar system 
constraints fairly easily is that, even in the vacuum spherically symmetric 
background, the Gauss-Bonnet scalar takes a non-vanishing value
$\G=12r_s^2/r^6$, where $r_s$ is the Schwarzschild radius.
In the solar system the GB term is much 
larger than the cosmological value $\G_* \sim H_*^4$.
Hence the inverse power-law terms such as $(\G_*^2/\G^2)^n$ 
($n>0$) are strongly suppressed even for the vacuum solution. 
This property is different from $f(R)$ gravity in which the 
Ricci scalar $R$ vanishes in the vacuum spherically symmetric 
background. In this case the presence of non-relativistic matter
with density $\rho_m$ leads to a non-vanishing Ricci scalar $R$
approximately proportional to $\rho_m$.
Due to the existence of matter, the local gravity gravity constraints 
can be satisfied for viable $f(R)$ models having the asymptotic 
behavior $f(R)=R-\lambda R_*\left[1-(R_*^2/R^2)^{n} \right]$
in the region $R^2 \gg R_*^2$.
In $f(\G)$ gravity we have shown that the contribution of 
matter density $\rho_m$ to the GB term ($\sim \rho_m^2$) 
can be negligible relative to the vacuum contribution ($\sim r_s^2/r^6$)
outside the area of Sun. 
Thus our analysis based on the vacuum spherically symmetric solution
is reliable to discuss the compatibility of $f(\G)$ models 
with solar system experiments.

\begin{acknowledgments}
The work of ADF is supported by the Belgian Federal Office for
Scientific, Technical and Cultural Affairs through the
Interuniversity Attraction Pole P6/11.
ST thanks financial support for JSPS (No.~30318802).
\end{acknowledgments}

\appendix
\section{Second-order equations for the model A}

In this Appendix we present second-order equations in terms of the 
expansion parameter $\ve$ for the model (\ref{modela}).
The second-order quantities $\psi_2$ and $\phi_2$
in the metric (\ref{permetric}) obey the following equations of motion
\begin{eqnarray}
& &( {\rho}^{4}-\rho-3\,{\rho}^{3}+3\,{\rho}^{2}) {\frac {\de\psi_2}
{\de\rho}}+ \left(3\,\rho -3\,{\rho}^{2}-1+{\rho
}^{3} \right) \psi_2+4096\,{\rho}^{10
}{\lambda}^{2}+ \left( 672\,\ln\rho+224\,\ln\ve -17488-112\,
\alpha \right) {\lambda}^{2}{\rho}^{9}\nonumber \\
& & + \left( -1040\,\ln\ve
+520\,\alpha-3120\,\ln 
\rho +29472 \right) {\lambda}^{2}{\rho}^{8}+ \left( -
24448+1776\,\ln\ve -888\,\alpha
+5328\,\ln\rho  \right) {\lambda}^{
2}{\rho}^{7} \nonumber \\
& & +\left( 664\,\alpha-3984\,\ln\rho+9952-1328\,\ln\ve 
\right) {\lambda}^{2}{\rho}^{6}+ \left(-1584-184\,\alpha
+368\,\ln\ve +1104\,\ln\rho
\right) {\lambda}^{2}{\rho}^{5}=0\,,
\label{appen1}
\end{eqnarray}
and 
\begin{eqnarray}
& & (-21\,{\rho}^{2}+21\,{\rho}^{3}+7\,\rho-7\,{\rho}^{4})\, 
{\frac {\de\phi_2}{\de\rho}} + \left( 7+7\,{\rho}^{
2}-14\,\rho \right)\phi_2+8512\,{\rho}^{10}{
\lambda}^{2} \nonumber \\
& & + \left( -34976+112\,\alpha-224\,\ln\ve
-672\,\ln\rho\right) {\lambda}^{2}{\rho}^{9}+ 
\left( 168\,\ln\rho-28\,\alpha+56\,\ln\ve
+56260 \right) {\lambda}^{2}{\rho}^{8} \nonumber \\
& &+\left( 1344\,
\ln\ve +4032\,\ln\rho -44440-
672\,\alpha \right) {\lambda}^{2}{\rho}^{7}
+ \left( -1960\,\ln\ve -5880\,\ln\rho+980\,\alpha+
17444 \right) {\lambda}^{2}{\rho}^{6} \nonumber \\
& &+ \left( 2352\,\ln\rho-392\,\alpha+784\,\ln\ve-2800 \right) 
{\lambda}^{2}{\rho}^{5}=0\,.
\label{appen2}
\end{eqnarray}
The solutions to this equation, discarding the homogeneous parts, are given by 
Eqs.~(\ref{psi2eq}) and (\ref{phi2eq}).


\end{document}